# Plasmonic- and Electronic-Enhancement-Free Coherent Raman Detection of Ångström-Scale Molecular Layers at Metal Interfaces

Toshiki Sugimoto[1,2,3]*, Tomoaki Ichii[1], Tsuneto Kanai[1], Ryu Yoshizawa[1,2], Shota Takahashi[1], Atsunori Sakurai[1,2,3], Keisuke Seto[1], and Jin Chengxiang[1,2]

[1] Institute for Molecular Science, National Institutes of Natural Sciences, Okazaki, Aichi 444-8585, Japan

[2] Graduate Institute for Advanced Studies, SOKENDAI, Okazaki, Aichi 444-8585, Japan

[3] Laser-driven Electron-acceleration Technology Group, RIKEN Spring-8 Center, Sayocho, Hyogo, 679-5148, Japan

KEYWORDS: interfacial Raman spectroscopy, coherent Raman spectroscopy, nonlinear spectroscopy, molecular vibration, metal interfaces

ABSTRACT: Coherent Raman scattering provides highly sensitive vibrational analysis through nonlinear light-matter interactions. However, its application to metal interfaces has remained challenging because the intrinsically large non-resonant background (NRB) of metals overwhelms weak interfacial molecular vibrational signals, making direct Raman detection without plasmonic or electronic enhancement highly challenging. Here, we report a time-frequency hybrid coherent Raman spectroscopy approach that overcomes this limitation and enables sensitive detection of ångström-thick molecular systems even on atomically flat metal surfaces. Our method employs a time-frequency engineered detection scheme that combines femtosecond pump and Stokes pulses with a time-delayed, asymmetrically shaped picosecond probe pulse. By exploiting instantaneous temporal response of the metal NRB, this pulse configuration effectively filters out the dominant metal NRB in the time domain while retaining a controlled residual NRB that acts as an internal local oscillator, enabling strong interferometric amplification of weak interfacial vibrational signatures. This all-optical coherent enhancement strategy establishes a new route for direct, non-invasive Raman detection of interfacial molecular systems across a wide range of surfaces without requiring structure- and material-specific plasmonic and electronic enhancement mechanisms.

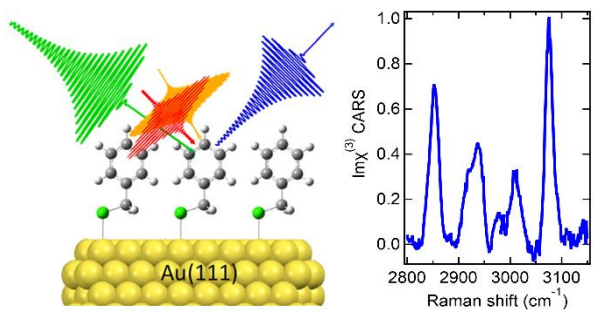

TOC

Molecular vibrations provide a direct fingerprint of chemical bonding, structure, and intermolecular interactions, making vibrational spectroscopy an indispensable tool across chemistry, physics, and materials science.[1-8] Raman spectroscopy, which probes vibrations through inelastic light scattering, has therefore become a cornerstone technique because it accesses infrared-inactive vibration, operates under ambient conditions, and offers broad chemical specificity for complex systems.[1,9-12] Despite these advantages, extending Raman spectroscopy to molecular systems confined at surfaces and interfaces remains fundamentally challenging, due to the extremely small number of molecules involved and intrinsically weak cross section of spontaneous Raman scattering.[13-18]

For interfacial molecular layers with thickness ranging from a few nanometers down to the ångström-scale, spontaneous Raman measurements typically require high excitation intensities, long acquisition times, and stringent experimental constraints to suppress fluorescence, while often yielding limited signal-to-noise ratios.[13-16] To overcome these limitations, , a wide range of enhancement strategies have been developed, including electronic resonance Raman spectroscopy,[2,11,19] and plasmonic field enhancement approaches such as surface-enhanced Raman spectroscopy (SERS),[11,17,18,20] shell-isolated nanoparticle-enhanced Raman spectroscopy (SHINERS),[11,21,22] and tip-enhanced Raman spectroscopy (TERS).[2,11,18,23-25] In some cases, charge-transfer interactions at metal-molecule interfaces provide an additional contribution to chemical enhancement.[11,17,18,26] While these approaches have enabled remarkable sensitivity gains and advances in interfacial Raman spectroscopy, their applicability and reliability often depend on specific nanostructures, materials, or molecular systems, and they may perturb interfacial molecular environment under investigation.[25,27] This has stimulated interest in complementary Raman methodologies that aim to achieve high sensitivity without relying on these enhancement strategies.

Coherent Raman scattering techniques, such as coherent anti-Stokes Raman scattering (CARS), offer an alternative route to highly sensitive vibrational spectroscopy by exploiting stimulated third-order nonlinear light-matter interactions.[1,28-32] In CARS, Raman active molecular vibrations with a frequency $\Omega_0$ are coherently driven by the frequency difference between pump ($\omega_1$) and Stokes ($\omega_2$) fields ($\omega_1 - \omega_2 \equiv \Omega = \Omega_0,$), and the resulting vibrational coherence is up-converted by a probe pulse ($\omega_3$) into an anti-Stokes signal at $\omega_{AS} = \omega_1 - \omega_2 + \omega_3$ through a four-wave mixing process. The coherent nature of this process typically yields directional signals that can be orders of magnitude stronger than spontaneous Raman scattering.[28-32] However, applying coherent Raman spectroscopy to interfacial molecular systems remains challenging because strong vibrationally non-resonant background (NRB) signals[33,32] arising from adjacent bulk media or substrates often overwhelm the weak vibrational response from interfaces. Various suppression strategies of the dominant NRB, such as polarization control detection,[34] interferometric detection,[35] and heterodyne detection,[36,37] have been developed to mitigate this problem, yet they remain insufficient for reliably detecting vibrational signals from interfacial layers.

For transparent dielectric substrates, where the vibrationally non-resonant third-order nonlinear susceptibility is typically modest,[28,38,39] specific reflection-based geometries and polarization schemes near the Brewster angle have been shown to suppress NRB contributions and enables coherent Raman detection of interfacial molecular layers.[33,40,41] In contrast, metal substrates pose a fundamentally more severe challenge. Owing to their strong optical reflectivity and large non-resonant third-order nonlinear susceptibility,[28,39] the intense NRB generated by metals cannot be effectively mitigated by optical geometry, preventing straightforward extension of coherent Raman techniques to metal-supported molecular layers.[42,43] As a practical consequence, coherent Raman spectroscopy of interfacial molecular systems on metal has often relied on plasmonic nanostructures,[18] thereby limiting generality and quantitative applicability.

In this work, we demonstrate a CARS spectroscopy approach that overcomes these limitations and enables sensitive detection of ångström-thick molecular layers on flat metal surfaces. Our method employs a hybrid time-frequency detection scheme,[32,44-49] in which femtosecond pump and Stokes pulses generate vibrational coherence, while a time-delayed, asymmetrically shaped picosecond probe pulse with a sharp femtosecond-scale rising edge selectively samples the nonlinear vibrational response[50-53] (Figure 1). By exploiting inherently instantaneous nature of the temporal response of the metal NRB, this engineered pulse configuration effectively filters out the dominant background in the time domain, achieving more than four orders of magnitude suppression. Then, rather than eliminating the NRB entirely, the controlled residual NRB is harnessed as an internal local oscillator, enabling coherent interferometric amplification of weak interfacial vibrational signals. This approach provides a direct and non-invasive route to highly sensitive Raman detection of ultrathin molecular layers even on atomically flat non-plasmonic metal substrates under electronically nonresonant conditions.

As a model system for metal-supported ångström-thick molecular layers free from plasmonic and electronic resonance enhancement, a self-assembled monolayer (SAM) of benzyl mercaptan (BM, ~$10^{14}$ molecules cm$^{-2}$)[54-56] was prepared on an atomically flat Au(111) surface. BM-SAM provides a structurally simple and chemically well-characterized interface, consists of a quasi-centrosymmetric phenyl ring (~ 5 Å thick) and an anti-centrosymmetric methylene unit (~ 2 Å thick), as illustrated in Figure 1(a). This structural motif naturally incorporates vibrational modes that either obey or violate the infrared-Raman mutual exclusion rule, offering an ideal platform for assessing the intrinsic sensitivity of the present CARS approach to Raman-active interfacial vibrations without relying on infrared activity.

All measurements were performed using near-infrared excitation, well away from the electronic resonances of both the molecular layer and the Au(111) substrate (below 540 nm). A three-color CARS scheme was implemented using a femtosecond Yb:KGW laser source, generating femtosecond pump ($\omega_1$, $\lambda_1$ = 1034 nm, FWHM ≈ 8.5 nm) and tunable broadband Stokes ($\omega_2$, $\lambda_2$ = 1150–1500 nm, FWHM ≈ 80 nm) pulses, together with a time-delayed, asymmetrically shaped picosecond narrowband probe pulse ($\omega_3$, $\lambda_3$ = 1034 nm, FWHM ≈ 0.7 nm (6.3 cm$^{-1}$). The probe pulse featured a sharp femtosecond rising edge followed by a picosecond decay, enabling selective temporal sampling of the vibrational response. These three beams were spatially overlapped on the sample in a reflection geometry with $p$-polarized excitation. In this configuration, the detected CARS signal in the *pppp* polarization combination is dominated by the *zzzz* component of the third-order nonlinear susceptibility tensor, $\chi^{(3)}_{zzzz}$, where $z$ is defined as the surface-normal direction. This dominance arises from the Fresnel local field factor at the metal surface.[17,57] No detectable thermal effects or laser-induced damage were observed under ambient conditions, confirming that the measurements were performed in a non-invasive regime under the present irradiation conditions.

To rationalize the design principle of the time-frequency hybrid three-color CARS approach, it is essential to consider the distinct temporal characteristics of vibrationally non-resonant and resonant nonlinear responses at metal-supported interfacial molecular systems. The non-resonant background (NRB) originates from an essentially instantaneous four-wave-mixing response of the metal substrate, whereas the vibrationally resonant CARS signal is governed by the free induction decay of molecular coherence persisting over the vibrational dephasing time $T_2$.[32,44-53] In conventional two-color CARS schemes, where the pump and probe fields are frequency-degenerate ($\omega_3 = \omega_1$), the complete temporal overlap of the pump, Stokes, and probe pulses inevitably generates a strong NRB, which often overwhelms weak vibrational responses from interfacial molecular layers. In contrast, the three-color CARS configuration ($\omega_3 \neq \omega_1$), combined with a time-asymmetric probe pulse with a femtosecond-scale sharp rising edge and a picosecond-scale duration, enables controlled temporal overlap with both the pump-Stokes excitation pulse pair ($\omega_1$ and $\omega_2$) and the subsequent vibrational free induction decay (Figure 1(b)). This temporal gating allows frequency-domain detection of the nonlinear optical response while maintaining high sensitivity to resonant vibrational contributions.

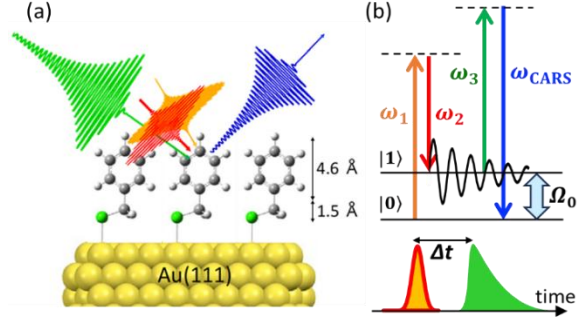

**Figure 1.** (a) Schematic experimental geometry of time-delayed femtosecond/picosecond hybrid three-colour CARS scheme applied to an ångström-scale self-assembled monolayer on atomically flat Au(111) substrate. (b) Energy-level diagram and temporal pulse sequence of the time-delayed three-color CARS scheme.

Figure 2(a) shows the time-delayed CARS intensity spectra, $I_{\text{CARS}}(\Omega; \Delta t)$, in the C-H stretching region as a function of Raman shift ($\Omega \equiv \omega_1 - \omega_2$) and probe delay $\Delta t$. At short probe delays ($\Delta t \lesssim 300$ fs), the vibrational resonant features are completely obscured by an intense NRB, $I_{\text{CARS}}(\Omega; \Delta t) \approx I_{\text{NRB}}^{(3)}(\Omega; \Delta t) \propto \left| \tilde{P}_{\text{NR}}^{(3)}(\Omega; \Delta t) \right|^2$, where $\tilde{P}_{\text{NR}}^{(3)}(\Omega; \Delta t)$ denotes the frequency-domain vibrationally nonresonant third-order polarization originating from the gold substrates. Because the vibrationally nonresonant third-order nonlinear susceptibility $\chi_{\text{NR}}^{(3)}$ is typically frequency independent,[28,31,32,46-49] the NRB exhibits a broadband, featureless spectral profile determined by the convolution of the spectral distributions of the pump and Stokes pulses. The decay profile of the NRB intensity with increasing probe delay $\Delta t$ is shown in Figure 2(b) and is well reproduced by numerical simulations that account for the asymmetric temporal profile of the probe pulse. Notably, as the probe delay $\Delta t$ increases, the NRB intensity is progressively suppressed by more than four orders of magnitude relative to that at zero probe delay ($\Delta t = 0$ fs).

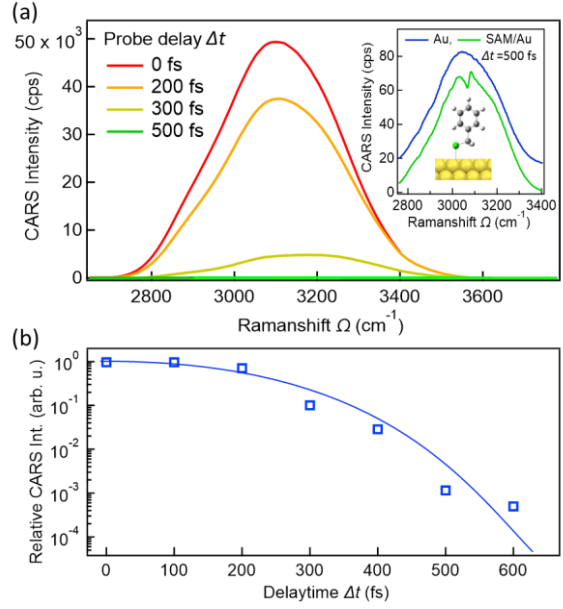

**Figure 2.** (a) Time-delayed CARS intensity spectra measured from Au(111) substrate covered with a benzyl mercaptan (BM) self-assembled monolayer (SAM) at various probe delay time $\Delta t$. Inset: CARS intensity spectrum at $\Delta t = 500$ fs (green), compared with that obtained from bare Au(111) under identical optical conditions (blue). These spectra are vertically offset for visual clearly. (b) Probe-delay-time dependence of the vibrationally nonresonant background (NRB) intensity, normalized to the value at $\Delta t = 0$ fs. The solid curve shows the simulated $\Delta t$ dependence of the NRB based on the temporal overlap of the pump, Stokes, and probe pulses.

As shown in the inset of Figure 2(a), a pronounced dip-like feature emerged in the broadband NRB spectrum at $\Delta t = 500$ fs. In contrast, this spectral feature is absent for a pristine Au(111) surface without molecular adsorption (inset of Figure 2(a)), confirming its molecular vibrational origin. The dip at ~3070 cm$^{-1}$ is very close to the resonant frequency of the highly symmetric phenyl C-H stretching modes that are Raman-active but infrared-inactive (Figure 4). Notably, this dip feature originates from the coherent superposition in the frequency domain between the vibrationally resonant third-order polarization, $\tilde{P}_R^{(3)}(\Omega; \Delta t)$, and the non-resonant polarization, $\tilde{P}_{NR}^{(3)}(\Omega; \Delta t)$. Under these conditions, the non-resonant contribution remains a few orders of magnitude larger than the resonant one (i.e., $\left|\tilde{P}_{NR}^{(3)}\right| \gg \left|\tilde{P}_R^{(3)}\right|$), so that the vibrational CARS signal is predominantly encoded in their interference term:

$$I_{CARS}(\Omega; \Delta t) \propto \left|\tilde{P}_{NR}^{(3)}(\Omega; \Delta t) + \tilde{P}_R^{(3)}(\Omega; \Delta t)\right|^2$$
$$\approx \left|\tilde{P}_{NR}^{(3)}(\Omega; \Delta t)\right|^2 + 2\text{Re}\left[\tilde{P}_{NR}^{(3)*}(\Omega; \Delta t)\tilde{P}_R^{(3)}(\Omega; \Delta t)\right], \quad (1)$$

while the purely resonant term $\left|\tilde{P}_R^{(3)}\right|^2$ is negligibly small. When the temporal width of the pump and Stokes pulses and the leading edge of the probe pulse are much shorter than the vibrational dephasing time $T_2$, and the probe bandwidth is narrower than the vibrational linewidth, the resonant polarization can be expressed as

$$\tilde{P}_R^{(3)}(\Omega;\Delta t) = \tilde{P}_{R0}^{(3)} \exp(-\Delta t/T_2)\chi_R^{(3)}(\Omega), \quad (2)$$

where $\chi_R^{(3)}(\Omega) = A_0/(\Omega_0 - \Omega - i/T_2)$ represents the vibrational resonant contribution to the third-order nonlinear susceptibility. This expression connects the time-domain decay of molecular coherence to the frequency-domain vibrational lineshape (Figure S4). Substituting Eq. (2) into Eq. (1) and introducing $\phi_{NR}$ as the phase of $\tilde{P}_{NR}^{(3)}$ gives

$$I_{\text{CARS}}(\Omega;\Delta t) \approx \left|\tilde{P}_{NR}^{(3)}(\Omega;\Delta t)\right|^2 + 2\left|\tilde{P}_{NR}^{(3)}(\Omega;\Delta t)\right|\tilde{P}_{R0}^{(3)} \exp(-\Delta t/T_2)\text{Re}\left[e^{-i\phi_{NR}}\chi_R^{(3)}(\Omega)\right]. \quad (3)$$

Normalizing the total CARS intensity $I_{\text{CARS}}(\Omega;\Delta t)$ by the pure NRB intensity $I_{\text{NRB}}(\Omega;\Delta t) \propto \left|\tilde{P}_{NR}^{(3)}(\Omega;\Delta t)\right|^2$ yields a simple form for the vibrational response:

$$\frac{I_{\text{CARS}}(\Omega;\Delta t)}{I_{\text{NRB}}(\Omega;\Delta t)} \approx 1 + S(\Delta t)\text{Re}\left[e^{-i\phi_{NR}}\chi_R^{(3)}(\Omega)\right]. \quad (4)$$

where $S(\Delta t) \equiv 2\tilde{P}_{R0}^{(3)}\exp(-\Delta t/T_2)/\left|\tilde{P}_{NR}^{(3)}(\Omega;\Delta t)\right|$. The spectral shape of the normalized CARS intensity is thus governed by the term $\text{Re}\left[e^{-i\phi_{NR}}\chi_R^{(3)}(\Omega)\right]$. When $\phi_{NR} = 0$ or $\pi$, vibrational response of the normalized CARS intensity follows $\text{Re}\chi_R^{(3)}(\Omega)$, resulting in a dispersive (derivative-like) line shape. In contrast, when $\phi_{NR} = \pm\pi/2$, the normalized CARS spectrum reflects $\text{Im}\chi_R^{(3)}(\Omega)$, leading to purely absorptive peak.

Figure 3(a) shows the normalized CARS spectrum obtained by dividing the measured CARS intensity spectrum by the NRB intensity spectrum at a probe delay of $\Delta t=$ 500 fs, based on the data shown in the inset of Figure 2(a). The normalized spectrum exhibits a pronounced derivative-like feature around ~3070 cm$^{-1}$, together with a weaker dispersive vibrational response near 3010 cm$^{-1}$ that originates from less symmetric phenyl C-H stretching modes that are both Raman- and infrared-active. As shown in Figure S7(c), these spectral features can be qualitatively reproduced by fitting with the two Lorentzian components, $\chi_R^{(3)}(\Omega) = \sum_{k=1,2} A_{0k}/(\Omega_{0k} - \Omega - i/T_{2k})$ with a nonresonant phase of $\phi_{NR} \approx \pi$. This analysis indicates that the vibrational contribution in the normalized CARS spectrum (Figure 3(a)) predominantly reflects the real part of $\chi_R^{(3)}(\Omega)$.

To further retrieve accurate spectral line shapes without assuming any prior model, the maximum entropy method (MEM)[31] was applied to the normalized CARS spectrum (Figure 3(a)). This approach successfully reconstructs both the real and imaginary components of $\chi_R^{(3)}(\Omega)$, providing a model-independent characterization of the interfacial vibrational response. The reconstructed $\text{Re}[\chi_R^{(3)}(\Omega)]$ and $\text{Im}[\chi_R^{(3)}(\Omega)]$ spectra obtained by the MEM analysis are shown in Figure 3(b). In the $\text{Im}[\chi_R^{(3)}(\Omega)]$ spectrum, clear vibrational peaks are observed at $\Omega \sim 3075$ cm$^{-1}$ and $\Omega \sim 3010$ cm$^{-1}$. Importantly, the $\text{Im}[\chi_R^{(3)}(\Omega)]$ spectrum directly corresponds to the line shape of the spontaneous Raman spectrum.[31] Therefore, the successful detection of these vibrational modes, arising from a molecular layer with ångström-scale thickness, clearly demonstrates the sensitivity of our coherent Raman spectroscopic approach for probing metal-supported ultrathin interfacial molecular systems without relying on plasmonic, electronic and chemical enhancement mechanisms.[18,58]

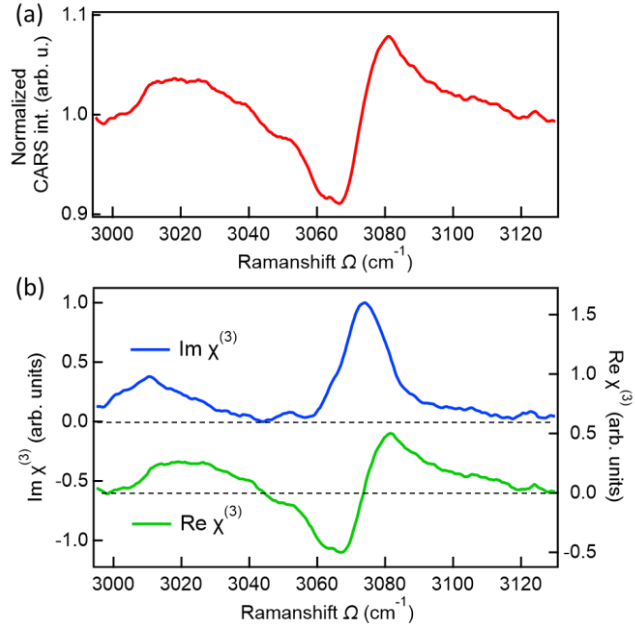

**Figure 3.** (a) Normalized CARS intensity spectrum measured at $\Delta t = 500$ fs in the C-H stretching region of the phenyl group of a BM-SAM on Au(111). (b) $\text{Im}\chi_R^{(3)}(\Omega)$ spectrum (blue, left axis) and $\text{Re}\chi_R^{(3)}(\Omega)$ spectrum (green, right axis) retrieved by the maximum entropy method from the normalized CARS intensity spectrum shown in (a).

By tuning the center wavelength of the $\omega_2$ Stokes pulse between 1150 and 1500 nm and performing similar time-delayed CARS measurements combined with MEM analyses at $\Delta t = 500$ fs, we obtained the $\text{Im}\chi_R^{(3)}(\Omega)$ and $\text{Re}\chi_R^{(3)}(\Omega)$ spectra covering both the fingerprint and C-H stretching regions (Figure 4). The $\text{Im}\chi_R^{(3)}(\Omega)$ spectra exhibit multiple distinct vibrational resonances (A-K), which can be classified into Raman-active but infrared-inactive modes (A-D, I and K) and modes that are active in both Raman scattering and infrared absorption (E-H, and J). As also shown in Figure S10 in more details, the Raman-active but infrared-inactive modes include highly symmetric phenyl-ring vibrations such as the ring-breathing (A), ring-bending (B), ring-bending coupled with $CH_2$ twisting (C), multiple ring-bending coupled with $CH_2$ twisting or wagging (D), phenyl C-H stretching vibration with minor ring-deformation character (I), and symmetric phenyl C-H stretching (K) modes. In contrast, Raman- and infrared-active modes are assigned as follows: (E) $CH_2$ wagging, (F), $CH_2$ symmetric C-H stretching involved in Fermi resonance with a $CH_2$ bending overtone, (G) phenyl C-H vibration coupled with ring-bending motion, (H) $CH_2$ bending overtone involved in Fermi resonance with the symmetric $CH_2$ C-H stretching vibration, and (J) multiple lower-symmetry phenyl C–H stretching modes.[55,56] The relatively enhanced experimental intensities of modes F and H compared with the calculations Raman intensities for an isolated BM molecule (middle of Figure 4(b)) may reflect adsorption-induced, mode-dependent modifications (chemical enhancement) of the Raman response at the metal interface.[26]

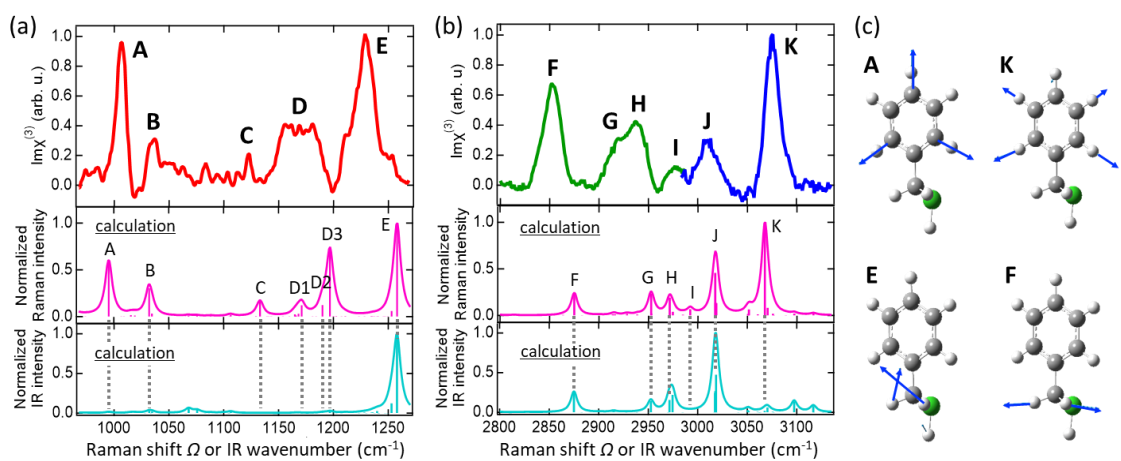

**Figure 4.** Vibrational spectra of the BM-SAM on Au(111) in (a) the fingerprint region and (b) the C-H stretching region. Top: experimentally retrieved $\mathrm{Im}\chi_R^{(3)}(\Omega)$ spectra. Middle and bottom: Raman intensity and IR absorption spectra of a BM molecule simulated by DFT calculation. (c) Representative vibrational motion of Raman-active but IR-inactive modes (A, K) of the quasi-centrosymmetric phenyl ring, and of modes that are both Raman- and IR-active (E. F) for the anti-centrosymmetric methylene group.

This systematic assignment reveals that the present interfacial CARS approach enables comprehensive probing of vibrational modes spanning a wide range of molecular symmetries, from highly symmetric Raman-dominant modes to lower-symmetry modes exhibiting mixed

Raman and infrared activity. We note that similar time-delayed CARS measurements have also been performed and the resultant $\text{Im}\chi_R^{(3)}(\Omega)$ spectra are obtained for other ultrathin molecular layers, indicating that the present approach is not specific to the BM-SAM shown here.

Building on these demonstrations of interfacial CARS molecular spectroscopy, we now address a fundamental question: whether coherent Raman techniques genuinely offer a sensitivity advantage over spontaneous Raman scattering for ångström-scale ultrathin interfacial systems, in contrast to bulk samples with micrometre-scale thickness.[29,30] From a general perspective, spontaneous Raman scattering is intrinsically incoherent and its signal scales linearly with the number of molecules $N$, while the intensity of CARS signals, derived from the purely resonant third-order polarization scale as $\left|\tilde{P}_R^{(3)}\right|^2 \propto N^2$. This quadratic scaling has led to the prevailing assumption that the intrinsic signal advantage of coherent Raman techniques diminishes in the ultrathin limit, particularly for molecular layers with ångström-scale thicknesses. Under pulse irradiation condition constrained by the damage threshold of the BM-SAM on Au, prior theoretical analyses[29,30] indicates that the time-averaged total photon flux generated by purely resonant CARS is comparable to, but typically a few times larger than, that of spontaneous Raman scattering for ångström-scale interfacial layers.

At this stage, however, it is crucial to emphasize that this comparison considers only the purely resonant contribution $\left|\tilde{P}_R^{(3)}(\Omega;\Delta t)\right|^2$ and does not yet account for a key feature of our interfacial CARS on metal substrates. In our case, the residual NRB signal from the metal substrate (inset of Figure 2(a)) serves as an intrinsic local oscillator[8,46,47,50] that coherently amplifies the weak interfacial vibrational response through optical interference. The vibrationally resonant component $\tilde{P}_R^{(3)}(\Omega;\Delta t)$, directly related to $\chi_R^{(3)}(\Omega)$, appears in the second term of Eq. (1). Since the interference term in Eq. (1) can be expressed as $2\left|\tilde{P}_{NR}^{(3)}(\Omega;\Delta t)\right|\left|\tilde{P}_R^{(3)}(\Omega;\Delta t)\right|\cos\Delta\Phi$, where $\Delta\Phi$ is phase difference between $\tilde{P}_R^{(3)}(\Omega;\Delta t)$ and $\tilde{P}_{NR}^{(3)}(\Omega;\Delta t)$, we define a coherent amplification factor, $w_R$, as the ratio of the amplified resonant term $2\left|\tilde{P}_{NR}^{(3)}(\Omega;\Delta t)\right|\left|\tilde{P}_R^{(3)}(\Omega;\Delta t)\right|$ to the intrinsic resonant term $\left|\tilde{P}_R^{(3)}(\Omega;\Delta t)\right|^2$ in the absence of NRB at $\Omega = \Omega_0$, $w_R \equiv 2\left|\tilde{P}_{NR}^{(3)}(\Omega_0;\Delta t)\right|\left|\tilde{P}_R^{(3)}(\Omega_0;\Delta t)\right|/\left|\tilde{P}_R^{(3)}(\Omega_0;\Delta t)\right|^2$. This can be equivalently rewritten as $w_R = 2\left|\tilde{P}_{NR}^{(3)}(\Omega_0;\Delta t)\right|^2/\left|\tilde{P}_{NR}^{(3)}(\Omega_0;\Delta t)\right|\left|\tilde{P}_R^{(3)}(\Omega_0;\Delta t)\right|$ and experimentally estimated to be $w_R \approx 28$ in our case (inset of Figure 2(a)). Combining this

interferometric amplification factor with the above estimate that the total photon flux generated by purely resonant CARS is a few times larger than that of spontaneous Raman scattering, the effective signal-generation efficiency of this interferometrically amplified interfacial CARS approach is expected to exceed that of spontaneous Raman scattering by approximately two orders of magnitude.

In addition to this signal-generation advantage, the overall sensitivity balance is further enhanced by signal-detection efficiency. Whereas spontaneous Raman photons are emitted isotropically and collected over a limited solid angle (typically a few %), coherent Raman signals are emitted directionally and collected with near-unity efficiency. Taking this signal directionality and the resulting collection efficiency into account, the experimentally detectable time-averaged photon flux of the present interfacial CARS approach is estimated to exceed that of spontaneous Raman scattering by approximately four orders of magnitude. In this context, we note that spontaneous Raman measurements performed under continuous-wave irradiation at 532 nm and 633 nm, using optical configurations and acquisition times comparable to those of the present CARS experiments, yielded no detectable signal from the BM-SAM on Au(111). This observation is consistent with earlier reports in which unenhanced spontaneous Raman signals from sub-nanometer adsorption layers on flat metal surfaces were only marginally detectable, even under substantially higher probe intensities and significantly longer acquisition times.[13-17]

The present approach enables direct access to Raman-active vibrational modes of ångström-scale metal-supported molecular systems. This capability is particularly significant for investigating chemically important yet IR-inactive interfacial molecules on metal surfaces, including homonuclear diatomic species such as $H_2$, $N_2$, and $O_2$, whose interfacial behavior has remained challenging to probe experimentally. The pulsed excitation scheme also provides a foundation for time-resolved coherent Raman studies of interfacial molecular dynamics, including IR-inactive high-frequency vibrations.

In summary, we have developed a coherent Raman spectroscopy approach capable of sensitively detecting ångström-scale ultrathin molecular systems supported on metal surfaces without requiring plasmonic and electronic enhancement. Our method employs a time-frequency engineered CARS scheme in which femtosecond pump and Stokes pulses are followed by a time-delayed, asymmetrically shaped picosecond probe pulse, enabling substantial suppression of the metal NRB. In this background-reduced regime, the residual NRB serves as an internal local oscillator, leading to coherent interferometric amplification of weak interfacial molecular vibrational signals. The directional emission of the CARS signal further enables efficient signal collection even with moderate numerical-aperture (NA) optics, facilitating practical measurements across a wide range of experimental geometries, in contrast to spontaneous Raman spectroscopy, which typically requires high-NA collection optics positioned close to the sample. Overall, our approach establishes a general optical strategy for high-sensitivity interfacial Raman spectroscopy with minimal structural and environmental constraints. By complementing structure- and material-specific plasmonic and electronic enhancement strategies, it provides a versatile platform for nanoscience, surface chemistry, and molecular interface research, where ultrathin interfacial molecular layers play central functional roles.

## ASSOCIATED CONTENT

**Corresponding Authors**

***Toshiki Sugimoto** — Institute for Molecular Science, National Institutes of Natural Sciences, Okazaki, Aichi 444-8585, Japan; Graduate Institute for Advanced Studies, SOKENDAI, Okazaki, Aichi 444-8585, Japan; Laser-Driven Electron-Acceleration Technology Group, RIKEN SPring-8 Center, Sayocho, Hyogo 679-5148, Japan
https://orcid.org/0000-0003-3453-6009
Email: toshiki-sugimoto@ims.ac.jp

**Authors**

**Tomoaki Ichii**— Institute for Molecular Science, National Institutes of Natural Sciences, Okazaki, Aichi 444-8585, Japan
https://orcid.org/0000-0002-7933-6134
**Tsuneto Kanai** — Institute for Molecular Science, National Institutes of Natural Sciences, Okazaki, Aichi 444-8585, Japan
https://orcid.org/0000-0003-1248-8122
**Ryu Yoshizawa**— Institute for Molecular Science, National Institutes of Natural Sciences, Okazaki, Aichi 444-8585, Japan; Graduate Institute for Advanced Studies, SOKENDAI, Okazaki, Aichi 444-8585
**Shota Takahashi** — Institute for Molecular Science, National Institutes of Natural Sciences, Okazaki, Aichi 444-8585, Japan
https://orcid.org/0000-0002-7191-5051
**Atsunori Sakurai** — Institute for Molecular Science, National Institutes of Natural Sciences, Okazaki, Aichi 444-8585, Japan; Graduate Institute for Advanced Studies, SOKENDAI, Okazaki, Aichi 444-8585, Japan; Laser-Driven Electron-Acceleration Technology Group,

RIKEN SPring-8 Center, Sayocho, Hyogo 679-5148, Japan

https://orcid.org/0000-0002-4410-7518

**Keisuke Seto** — Institute for Molecular Science, National Institutes of Natural Sciences, Okazaki, Aichi 444-8585, Japan

https://orcid.org/0000-0002-2859-6593

**Jin Cheng Xiang**— Institute for Molecular Science, National Institutes of Natural Sciences, Okazaki, Aichi 444-8585, Japan; Graduate Institute for Advanced Studies, SOKENDAI, Okazaki, Aichi 444-8585

**Author Contributions**

T.S. coordinated the project and designed the research; T.I., T.S., A.S., R.Y., and K.S. developed the experimental setup; S.T., T.I. R.Y., and J.C. prepared the samples,; T.I., R.Y., and K. S. performed the experiments; T.I. T.K., R.Y., S.T., and K.S. analyzed the data; S.T. carried out the DFT calculation; All authors discussed the results; T.S. wrote the manuscript with assistance from T.I. T.K. ST. and A.S.

**Notes**

The authors declare no competing financial interests.

**Acknowledgements**

This study was supported by JSPS KAKENHI Grant-in-Aid for Scientific Research (A) [22H00296], Grant-in-Aid for Challenging Research (Exploratory) [21K18896, 24K21759], Grant-in-Aid for Scientific Research (B) [23K26548], ATLA Japan Innovative Science and Technology Initiative for Security [JPJ004596], JST-FOREST [JPMJFR221U], JST-CREST [JPMJCR22L2], JST-K Program, [JPMJKP24W1], Amada Foundation General Research and Development Grant [AF-2021212-B2], Joint Research by the National Institutes of Natural Sciences (NINS) [01112104], and by the Special Project by Institute for Molecular Science [IMS programme 22IMS1101].